\pdfoutput=1 
\documentclass[12pt]{article}

\usepackage{inputenc}

\usepackage{amsmath,amssymb,bbm,color}
\usepackage{amsbsy}
\usepackage{fixmath}
\usepackage{euscript}
\usepackage{graphicx}
\usepackage{cite}
\numberwithin{equation}{section}

\usepackage{epsf} 

\newcommand{\codeMark}[1]{{\it #1}} 
\newcommand{\commandMark}[1]{{\it #1}} 
\newcommand{\classMark}[1]{{\it #1}} 

\newcommand{\dirFileMark}[1]{{\it #1}}

\begin{document}

\begin{titlepage}

\vspace{5mm}
\begin{center}
    {\Large\bf Software for calculations of the Higgs boson lineshape\\ in future lepton colliders}
\end{center}

\vskip 10mm
\begin{center}
{\large S.\ Jadach$^a$ and R.A.\ Kycia$^b$}

\vskip 2mm
{\em $^a$The Henryk Niewodnicza\'nski Institute of Nuclear Physics,\\ 
Polish Academy of Sciences,\\
  ul.\ Radzikowskiego 152, 31-342 Krak\'ow, Poland}\\
\vspace{1mm}
{\em $^b$T. Ko\'sciuszko Cracow University of Technology, \\ 
Faculty of Physics, Mathematics and Computer Science, \\ 
ul. Warszawska 24, 31-155 Krak\'ow, Poland}\\
\end{center}

\vspace{5mm}
\begin{abstract}
\noindent
Simple software for Monte Carlo(MC) calculation of the Higgs boson lineshape and its beam broadening effect by the beam energy dispersion was described. The software bases on FOAM \cite{foam:1999} adaptive MC integrator from ROOT library \cite{Root}. The software is intended for those who want to reproduce results from paper \cite{JadachKycia} and experiment with different parameters  for the lineshapes and QCD correction factors. Parallel version of the software based on MPI \cite{MPI} is also described.
\end{abstract}

\vspace{2mm}
\begin{description}
\item{\em Keywords:} QCD, the Higgs boson cross section, energy scan, Monte Carlo, FOAM, beam spread, initial radiation state.
\end{description}

\vspace{15mm}


\end{titlepage}

\newpage
\tableofcontents

\newpage
\section{Introduction}

The software presented here was used to produce results from paper 
\cite{JadachKycia} and therefore it is advisable to study the paper before start to experiment with the programs. In \cite{JadachKycia} there is also all theory required to understand the output. The software is available at \cite{DownloadSoftware}.

This paper is organized as follows. In the next section requirements for running the software is provided. Then a short example follows. Next, general idea of the software design is described. Finally a short introduction to the parallel version based on MPI is presented for those who want to experiment with parallel computation.

\section{Requirements}
The software requires rather standard and free software available in most Unix-like systems.
In order to compile and run programs mandatory tools are as follows
\begin{itemize}
 \item{Unix-like system - POSIX compatible, e.g., Linux;}
 \item{Make compatible system, e.g., GNU make \cite{GNUMake};}
 \item{g++ compiler from GNU Compiler Collection \cite{GCC};}
 \item{ROOT library \cite{Root} \footnote{For ROOT 6 or higher version it is compiled using C++ 2011 standard. In this case, when GNU \codeMark{g++} compiler is used then the flag \commandMark{-std=c++11} should be added during compilation for compatibility, see Makefile}.}
\end{itemize}
There is also additional software that is required to perform more advanced operations
\begin{itemize}
 \item{Doxygen \cite{Doxygen} (optional) if documentation from the code is required to generate;}
 \item{Valgrind \cite{Valgrind} (optional) if advanced debugging is required;}
 \item{MPI \cite{MPI}, e.g., OpenMPI (optional) if the user wants to experiment with parallel version of the program;}
\end{itemize}

\section{1 minute example}
Simplified version of the program was constructed to familiarize new users with its   philosophy of design.

\begin{figure}[h!]
  \centering
    \includegraphics[width=0.5\textwidth]{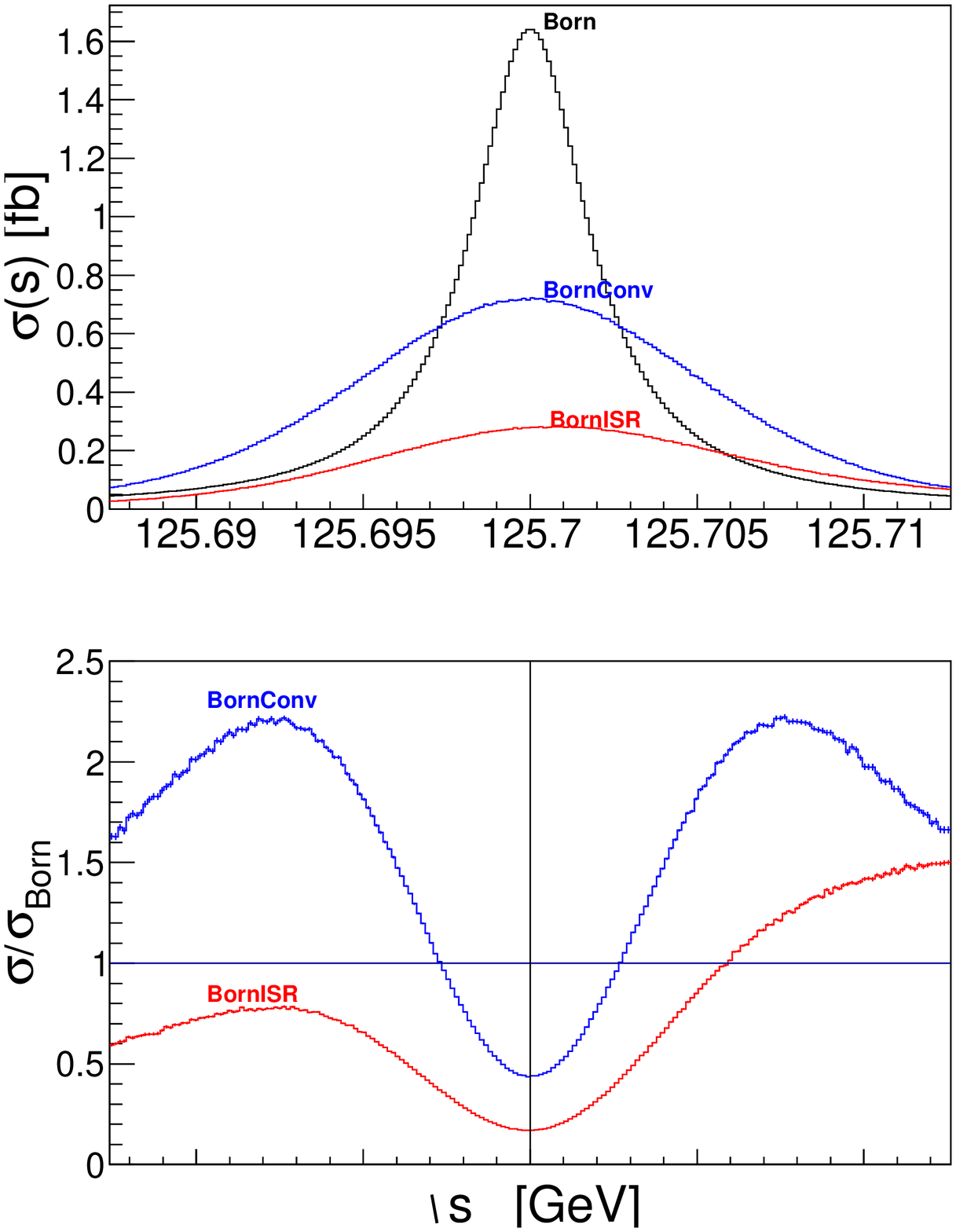}
    \caption{In the upper panel of the plot there is a Born term for the cross section of $ee\rightarrow H$ reaction, the Born term convoluted with the Gaussian distribution for the energy spread $\delta_{E}=4.2MeV$ and similar plot with ISR corrections included. The lower panel plot shows ratios of the last two plots by the Born term.}
    \label{Fig:SimplePlotEPS}
\end{figure}

In order to run the program go to \dirFileMark{Simple} directory and then type in the console \commandMark{make run}. When the program run ends, which should take no more than 1 minute, you will be able to see the example plot in \dirFileMark{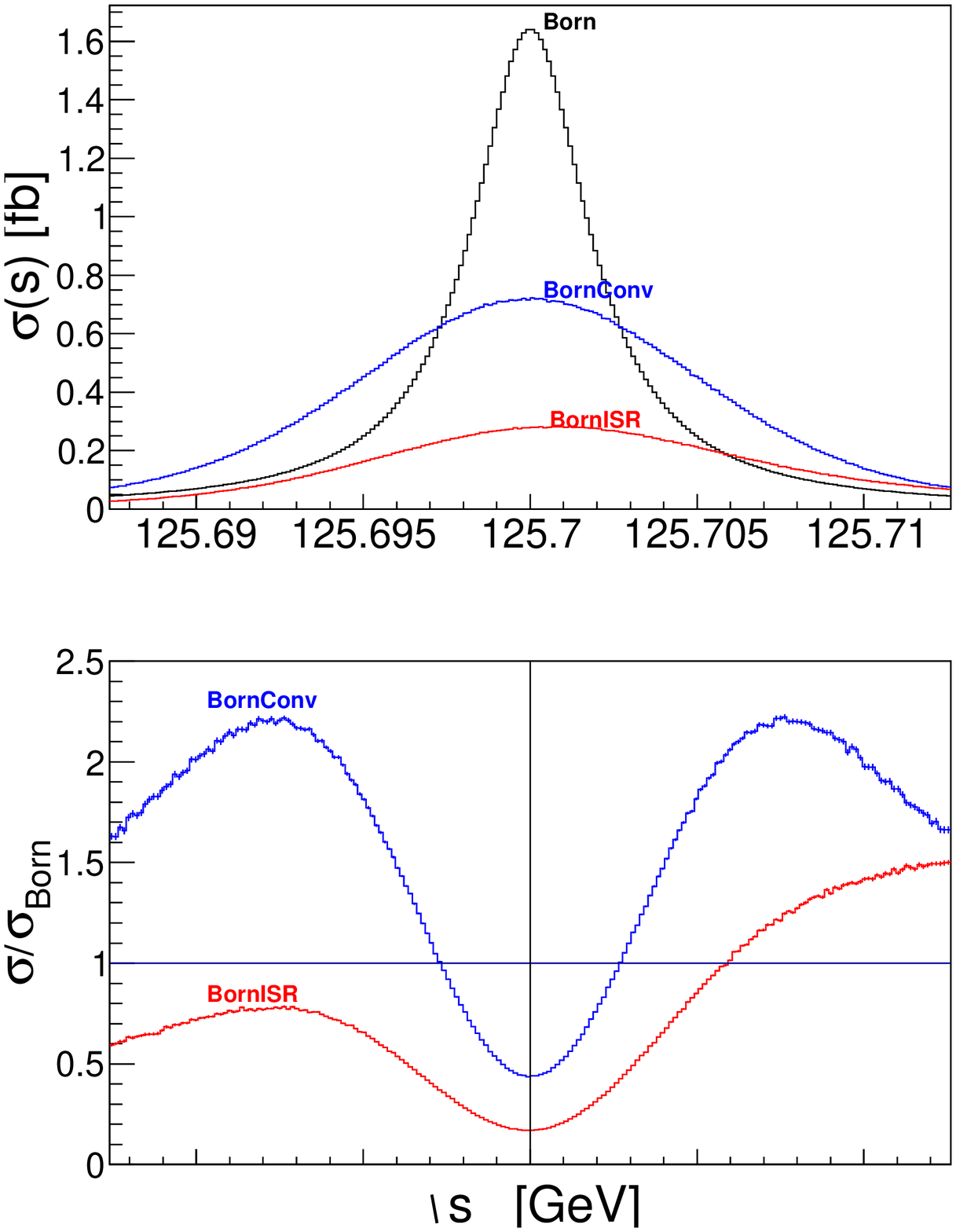} file, see Fig. \ref{Fig:SimplePlotEPS}.
These plots are only a small number of capabilities of the library of functions in the full program. 

Now we will analyze the code. Please open the \dirFileMark{main.cxx} file in your favorite editor. It is good to choose such an editor that has support for C/C++ programming. Let us start analysis of the code - scroll down the code and stop at the beginning of the \codeMark{main()} function. The first relevant instruction is
\begin{verbatim}
 long NevTot = 10000000;
\end{verbatim}
which sets up the statistic for Monte Carlo integrator. The error of integration for ISR corrections or convolution with beam center energy spread is roughly proportional to $\frac{1}{\sqrt{NevTot}}$. On the other hand the time of integration is proportional to the statistics. 

In the next lines, histograms are created. The line
\begin{verbatim}
 MakeBorn(string("BornH") );
\end{verbatim}
creates the Born term histogram and saves it in \dirFileMark{BornH.root} file. Then in the line
\begin{verbatim}
 MakeConvBorn( string("histo-sig04-born"), 0.0042, NevTot );
\end{verbatim}
the histogram for Born term convoluted with the Gaussian distribution for the energy spread $\delta_{E}=0.0042GeV$ is saved in the file \dirFileMark{histo-sig04-born.root}. Finally, the similar plot including ISR corrections is created in 
\begin{verbatim}
 MakeISR( string("histo-sig04-isr2"), 2, 3, 0.0042, NevTot );
\end{verbatim}
and it is saved in \dirFileMark{histo-sig04-isr2.root} file.

Now we can create plot in the function
\begin{verbatim}
 plotSimpleFigs();
\end{verbatim}
This function is customized by the user to adjust the plots to the user needs. We will analyze main parts of the function. Please scroll the code to \codeMark{plotSimpleFigs()} function. The first part
\begin{verbatim}
TFile DiskFileBorn(  "./histo-sig04-born.root"); 
TFile DiskFileISR2(  "./histo-sig04-isr2.root"); 
TFile DiskFileBornH(  "./BornH.root");   

TH1D *h_Born        = (TH1D*)DiskFileBornH.Get("h_SigEne");
TH1D *h_SigEneBorn  = (TH1D*)DiskFileBorn.Get("h_SigEne");
TH1D *h_SigEneISR2  = (TH1D*)DiskFileISR2.Get("h_SigEne");
\end{verbatim}
is responsible for retrieving histograms from the root files. Next, the values of the Higss mass(\codeMark{m\_MH}) and width (\codeMark{m\_GamH}) is retried from the \classMark{TDesity} object 
\begin{verbatim}
TDensity* Density = new TDensity();
double MH   = Density->m_MH;
double GamH = Density->m_GamH;
\end{verbatim}
The class \classMark{TDensity} is contained in an appropriate header and class implementations files. This class contains integrand function and all relevant physical constants. 

In the next part ROOT specific operations are performed in order to format plots.

\begin{figure}[h!]
  \centering
    \includegraphics[width=0.5\textwidth]{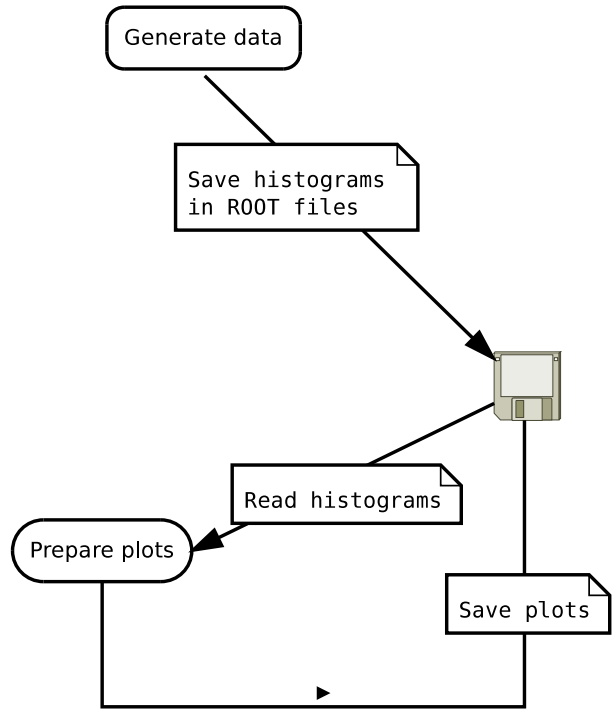}
    \caption{First histograms are created and saved in root files on disk. Then they are retrieved, adjusted and saved on disk in the post script format.}
    \label{Fig:Flowchart}
\end{figure}

The same idea on the large scale is used in the full software - theres is more cases for data generation and more plot functions that prepare customized plots; Fig. \ref{Fig:Flowchart} describes general concept.

\section{General overview of the program}
In this section we give description of general software from \dirFileMark{Full} directory. 

The software contains a few parts:
\begin{itemize}
 \item {\dirFileMark{Makefile} compatible with GNU make;}
 \item {Class \classMark{TDensity} in files \dirFileMark{TDensity.h} and \dirFileMark{TDensity.cxx} that contains integrands for Born convolution and ISR contribution. It inherits from \codeMark{TFoamIntegrand} class as it is required by FOAM \cite{foam:2002, foam:1999};}
 \item {Main program and function library in \dirFileMark{main.cxx} that contains functions described below and \codeMark{main()} function.}
\end{itemize}
The following commands are defined in \dirFileMark{Makefile}:
\begin{itemize}
 \item {\commandMark{make run} - compile and run program;}
 \item {\commandMark{make clean} - clean executables;}
 \item {\commandMark{make cleanest} - clean executables, root, pdf and eps files;}
 \item {\commandMark{make Generate-doc} - generate documentation from the code and display HTML version in Firefox web browser;}
\end{itemize}
Makefile is set up for parallel compilation on maximal number of cores available on the computer.

FOAM \cite{foam:2002, foam:1999} adaptive integrator requires the density distribution. The class \classMark{TDensity} implements density distribution for the sole Born distribution, the Born distribution with convolution with the Gaussian distribution and with additional ISR corrections \cite{JadachKycia}. The class can calculate cross section for electron and muon Higgs production. The type of reaction can be switched by (un)commenting appropriate \codeMark{define} statement in \dirFileMark{TDensity.h} file, e.g., default setup for $ee\rightarrow H$ reaction is as follows
\begin{verbatim}
#define ELECTRON
//#define MUON 
\end{verbatim}

The class \classMark{TDensity} contains fields which are responsible for selecting appropriate distributions:
\begin{itemize}
 \item{\codeMark{m\_ISROn} - $0$ - ISR corection is OFF, only the Born shape; $1$ - ISR correction ON;}
 \item {\codeMark{m\_keyISR} - selects ISR type of \cite{JadachKycia}: $0$ -(a); $1$- (b); $2$- (c);}
 \item {\codeMark{m\_kDim} - dimension of distribution to integrate. There are the following combinations possible:}
 \begin{itemize}
  \item {If \codeMark{m\_ISROn=0} then \codeMark{m\_kDim = 2} - convolution of the Gauss distribution with Born.}
  \item {If \codeMark{m\_ISROn=1} then \codeMark{m\_kDim = 2} - machine energy spread OFF; 3 - machine energy spread ON; }
 \end{itemize}
 \item {\codeMark{m\_sigE} - energy spread in $GeV$ if applicable(\codeMark{kDim = 3}).}
\end{itemize}

Let us give a few examples of combinations:
\begin{itemize}
 \item {\codeMark{m\_ISROn = 0, m\_kDim = 2, m\_sigE = 0.0042} - Born convoluted with the Gaussian distribution for the energy spread $\delta_{E}=4.2MeV$;}
 \item {\codeMark{m\_ISROn = 1, m\_kDim = 2, m\_keyISR=2} - Born with ISR (c);}
 \item {\codeMark{m\_ISROn = 1, m\_kDim = 3, m\_keyISR=2, m\_sigE = 0.0042} - Born with ISR (c) convoluted with the Gaussian distribution for the energy spread $\delta_{E}=4.2MeV$;}
\end{itemize}
However, there is no necessity to remember these combinations as there are available functions that set up all legitimate combinations. They were used in the simplified example and described below.

Let us now focus on \dirFileMark{main.cxx} file. The file contains two types of functions. The first kind of functions generates cross section distribution and save them into root file. These functions set up appropriately \classMark{TDesity} object, and use  FOAM to integrate distributions, create histograms and save them into root file. This root file contains two histograms
\begin{itemize}
 \item {h\_Ene - contains histograms of cross section as a function of energy;}
 \item {h\_NORM - contains two bins: the first one at $0.5$ that contains luminosity, and the second one at $1.5$ that contains number of events in h\_Ene histogram.}
\end{itemize}
The functions are as follows. The first function create Born term histogram and save it in a root file
\begin{verbatim}
void MakeBorn( string filename = "BornH" )
\end{verbatim}
The second function makes Born term convolution with the Gauss distribution and save it into root file
\begin{verbatim}
 void MakeConvBorn( string filename = "histo1", 
 Double_t sigE = 0.0041, long NevTot =  1000000 ) 
\end{verbatim}
The last important 'production' function makes Bron term with ISR contribution convoluted with the Gaussian distribution and save it into root file
\begin{verbatim}
 void MakeISR( string filename = "histo1", Int_t keyISR = 2,
 Int_t kDim = 3, Double_t sigE = 0.0041, long NevTot =  1000000 )
\end{verbatim}

The second kind of functions retrieve these histograms, do formating and save histograms into EPS files. These includes
\begin{verbatim}
 void plotISRabc( void )
\end{verbatim}
which prepares plots for three types of ISR contributions and save it in EPS file \dirFileMark{ISRabc.eps}. The next one 
\begin{verbatim}
 void plotISR123( void )
\end{verbatim}
prepares plots with ISR contribution and its convolution for $\sigma_{E} = 4.2MeV$ and $8MeV$. Plots are saved in \dirFileMark{ISR123.eps}. Next two functions
\begin{verbatim}
void plotBorndelta( void )
void plotISRdelta( void )
\end{verbatim}
prepare Born and ISR convolution plots with the Gauss distributions for different beam energy spread values. The last two functions are responsible for creating and plotting cross section dependence on the beam energy spread value
\begin{verbatim}
 int makeISRsigEDistribution( int nbins = 100, long NevTot =  1000000 )
 int plotISRsigEDistribution( int nbins = 100, long NevTot =  1000000 )
\end{verbatim}
These plot functions prepare plots that can be seen in paper \cite{JadachKycia}.

\section{Parallelization with MPI}

In this section description of parallel version of programs will be provided. It can be interesting for those who wants experiment with this idea. They are located in \dirFileMark{MPI} directory.

The processes of generation of histograms are separate, independent and require large amount of time for large statistic. Therefore, they are ideally suitable for parallelization. There are different approaches to the issue. Our approach is not very sophisticated one, however, conceptually simple. It relies on MPI(Message Passing Interface)\cite{MPI}, specifically OpenMPI technology for C/C++. MPI creates software infrastructure for communication between processes by MPI library calls. It is popular standard in High Performance Computing. Process management is implemented using the well known producer consumer design pattern \cite{Mordehai}. The number of processed used in computations is controlled by \codeMark{NOP} variable in \dirFileMark{Makefile}.

There are two programs. The first one in the subdirectory \dirFileMark{Basic} generates plots for all types of cross sections and the second one in the subdirectory \dirFileMark{sigEPlots}, which generates plots for dependence of cross section on the center of mass beam energy spread.

\begin{figure}[h!]
  \centering
    \includegraphics[width=0.3\textwidth]{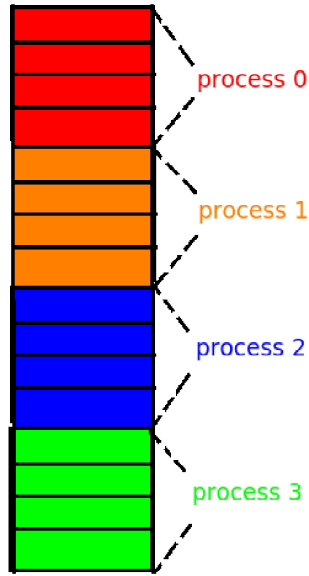}
    \caption{Data distribution of the array among the processes.}
    \label{Fig:ArrayDistribution}
\end{figure}

\begin{figure}[h!]
  \centering
    \includegraphics[width=0.5\textwidth]{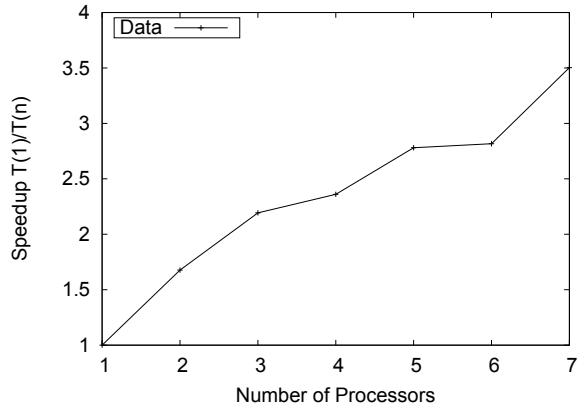}
    \caption{Speedup for \dirFileMark{Basic} program. It was calculated on Intel(R) Core(TM) i7-3610QM CPU @ 2.30GHz processor. Each process worked on single core and as there was only $8$ cores therefore there is inefficient to run more than $7$ processes at the same time. Here $T(i)$ is the time of computation for $i$ processes.}
    \label{Fig:Speedup}
\end{figure}

In the first program, the histograms are created according to the data in the array \codeMark{data}. Every cell of this array is of the struct type \codeMark{genData} which contains parameters of generation, e.g., beam energy spread. Then every process depending on its unique identification number generates histograms for its range of data in the array, see Fig. \ref{Fig:ArrayDistribution}.
The starting and ending index for every process can be calculated by rather standard functions 
\begin{verbatim}
int startIndex( int N, int workers, int rank);
int stopIndex( int N, int workers, int rank);
\end{verbatim}
where \codeMark{N} is the dimension of the array, \codeMark{workers} is the number of all processes and \codeMark{rank} is the unique rank of the process. Simple check for the speedup for the first program is presented in Fig. \ref{Fig:Speedup}.

In the second program every process generates the values of cross section for given range of values of beam energy spread from the whole energy spread range. The value of energy spread for the bin centers are stored in shared array - every process calculates cross section for given range of energy spread at given beam energies. These computed values are stored by processes in the arrays shared by processes. The values are gathered by the main process. This process then generates plots. In this program it is important that the size of the arrays shared between processes has to have multiplicity of the number of processes used in computation. Here the speedup of calculations is significant. However as the number of computations for process is the multiplicity of the number of processes therefore there is no simple measure of this speedup.

In the implementation the static scheduling of tasks was used, which is the simplest and the most ineffective type of scheduling. Therefore there is a big field for improvements.

\section{Conclusions}
Description of library and program for calculations of beam energy spread influence on cross section lineshape of Higgs boson was provided. Structure of the program can be easily adapted to other similar calculations. In the end simple parallelization of the program was described.

\providecommand{\href}[2]{#2}

\end{document}